\documentclass{article}

\usepackage{natbib}
\usepackage{mathptmx}
\usepackage{helvet}
\usepackage{courier}
\usepackage{graphicx}
\usepackage{multicol}
\usepackage[bottom]{footmisc}
\usepackage{aas_macros}
\usepackage{amsmath}
\usepackage{hyperref}
\usepackage{setspace}

\def\araa{ARA\&A}
\def\apj{ApJ}
\def\apjl{ApJL}
\def\apjs{ApJS}
\def\ao{Appl.~Opt.}

\def\aap{A\&A}

\def\mnras{MNRAS}

\begin{document}
\noindent {\bf{\Large Radial Migration in Spiral Galaxies}}\\

\noindent {\large Lessons from the Local Group, Seychelles, May 19-23, 2014}\\

\noindent {\small Rok Ro\v{s}kar, Institute for Computational Science, University of Z\"urich, 
Winterthurerstrasse 190, CH-8057 Z\"urich, Switzerland}\footnote{roskar@physik.uzh.ch}
\\
\small{Victor P. Debattista, Jeremiah Horrocks Institute, University
  of Central Lancashire, Preston, PR1 2HE, UK}

\bigskip

\begin{abstract}

The redistribution of stars in galactic disks is an important aspect
of disk galaxy evolution. Stars that efficiently migrate in such a way
that does not also appreciably heat their orbits can drastically
affect the stellar populations observed today and therefore influence
constraints derived from such observations. Unfortunately, while the
theoretical understanding of the migration process is becoming
increasingly robust, there are currently few specific observable
predictions. As a result, we do not yet have a clear handle on whether
the process has been important for the Milky Way in the past or how to
constrain it. I discuss some of the expected qualitative outcomes of
migration as well as some current controversies.

\end{abstract}

\section{Introduction}

Galactic archaeology attempts to infer past history of our Milky Way
disk from the rich datasets detailing stellar populations in terms of
their chemistry, kinematics and ages. However, in order to ease
interpretation, ``population'' is often loosely used to imply
membership based on \emph{current} cohabitation of a similar part of
the Galaxy. This is borne out in the use of terminology like ``the
Solar Neighborhood'' and the characterization of galactic disk
properties in terms of the one-dimensional distance from the galactic
center. For the most part, such groupings seem to make sense. After
all, if the guiding radius $R_g$ is defined by a star's angular
momentum, i.e. $R_g \sim L_z/V_c$ where $L_z$ is the $z$ component of
the angular momentum and $V_c$ is the circular velocity, then the
deviations from this radius during its orbit due to excess orbital
energy are constrained to $< 2$~kpc \emph{for the oldest stars} near
the sun \citep{Binney:2007}. Furthermore assuming that the disk is
mostly axisymmetric, then viewing the Solar Neighborhood as a
representative random sample around $\sim R_{\odot}$ and using it to
infer past history at this radius seems reasonable.

This simple assumptions is remarkably powerful. By combining our
knowledge of stellar evolution and metal production, just the
metallicity distribution function (MDF) and the age-metallicity
relation (AMR) are needed to infer the entire history of (this part
of) the MW disk. Unfortunately, in its simplest incarnation, this type
of modeling fails spectacularly in what is largely known as the
'G-dwarf problem', because it over-predicts the relative number of
metal-poor stars (e.g. \citealt{Searle:1972, Tinsley:1975}). One of
the explanations for too many metal-poor stars is that the available
gas is depleted too soon and therefore the simplest solution is to
allow for an inflow of gas from the outside \citep{Larson:1974,
  Lynden-Bell:1975}. In this way, it is recognized that of course the
Solar Neighborhood does not exist in isolation from other parts of the
Galaxy, but that it is only a part of the whole.

Although gas is regarded free to stream in and out of the volume
defining a local patch of the disk, stars are not usually assumed to
do so due to small radial orbital excursions described above. While
the authors themselves cautioned about observational bias resulting in
the AMR, \citet{Edvardsson:1993} data has frequently been used to
support the idea that local stars of different ages represented
generations of stars born out of the same recycled gas. This fits well
with most one-dimensional chemical evolution models
(e.g. \citealt{Prantzos:1995}), which robustly predict a strong
age-metallicity relation. However, more recent data (e.g. the
Geneva-Copenhagen survey \citealt{Holmberg:2009} and its various
re-interpretations, and the Gaia-ESO survey \citealt{Bergemann:2014})
show instead that the AMR is flat, that is, the AMR does not exist
(\citealt{Bergemann:2014} do note, however, the lack of
solar-metallicity stars older than 10 Gyr). The most striking feature
of these observations is the large scatter in [Fe/H] at each age,
which cannot easily be explained away by asymmetric drift
arguments. These observations therefore expose a need to rethink disk
evolution and let the stars jump out of the box.

\section{Stellar Migration}

If stars from very different parts of the disk can be found together
today, and if there is a metallicity gradient in the star-forming gas,
then a large spread in metallicity at each stellar age is
expected. Consequently, much like gas flows were the answer to the
G-dwarf problem, radial mixing of stars presents a convenient solution
for the dispersed AMR problem described above. The idea that stars
should be diffusing through the disk was first proposed by
\citet{Wielen:1977}, who suggested that determining the birth places
of stars may be complicated due to orbital diffusion. However, because
only diffusion due to random scattering off irregularities like GMCs
was considered, the guiding radii $R_g$ were shown to stay nearly
constant \citep{Grenon:1987}. Nevertheless, \citet{Wielen:1996}
postulated that he Sun itself must have moved away from its birthplace
in the inner disk, based on the fact that its metallicity is high
compared to other stars of similar age in its vicinity. Using
simplified arguments about the evolution of the Milky Way metallicity
gradient, they constrained the Sun's birthplace to be $\sim2$~kpc
closer to the Galactic center.

However, \citet{Sellwood:2002} showed that the treatment used in
\citet{Wielen:1977}, which considered only random scattering off GMCs,
was insufficient as it did not take into account resonant interactions
with spiral arms. In particular, \citet{Sellwood:2002} showed that if
a star's orbital speed matches the speed of a passing transient spiral
wave, i.e. the star is at the corotation resonance of the spiral, the
star's angular momentum $L_z$ can be altered on a very short
timescale. Due to the conservation of the Jacobi integral in such an
interaction, the changes in energy and angular momentum are related
simply by
\begin{equation}
\Delta E = \Delta L_z \Omega_p,
\end{equation}
where $\Omega_p$ is the pattern speed of the passing spiral
wave. \citet{Sellwood:2002} showed also that the radial action, $J_R$
of the migrating star changes as
\begin{equation}
\Delta J_R  = \frac{\Omega_p - \Omega}{\omega_R} \Delta L_z,
\end{equation}
where $\Omega$ and $\omega_R$ are the azimuthal and radial frequencies
of the star. Therefore, at corotation, angular momentum and energy are
exchanged without causing additional heating to the orbit, in stark
contrast to changes in $L_z$ at the Lindblad resonances where stars
heat very efficiently. This remarkable result presents a serious
hurdle for efforts trying to reconstruct Galactic history from
present-day observations. Stellar kinematics are typically used to try
and infer the dynamical histories of stars, i.e. postulate whether
they had heated, were accreted, had interacted with the bar etc. If
stars may change their guiding radii by \emph{several} kpc without
acquiring any kinematical trace of the process, it becomes very
difficult to disentangle which stars are actually telling the story of
this part of the Galaxy, and which are part of another history
altogether.

There is an important caveat with regard to the corotation
resonance. If the perturbing pattern is steady, the corotation
resonance results simply in orbit trapping. In this case, a star on
the inside of corotation gains angular momentum passing to the outside
of corotation, where it is now leading the pattern slightly and being
pulled back by the overdensity loses the previously gained angular
momentum. This trapping results in a horseshoe orbit which in the
steady-state remains at a constant $L_z$. Therefore, for the
corotation resonance to impart secular changes to a star's angular
momentum, the pattern must be transient but last just long enough to
deposit the star on the other side of the resonance before it gets
pulled back in the other direction.

\section{Stellar migration mechanisms in simulations}
\begin{figure*}
\centering
\includegraphics[width=\textwidth]{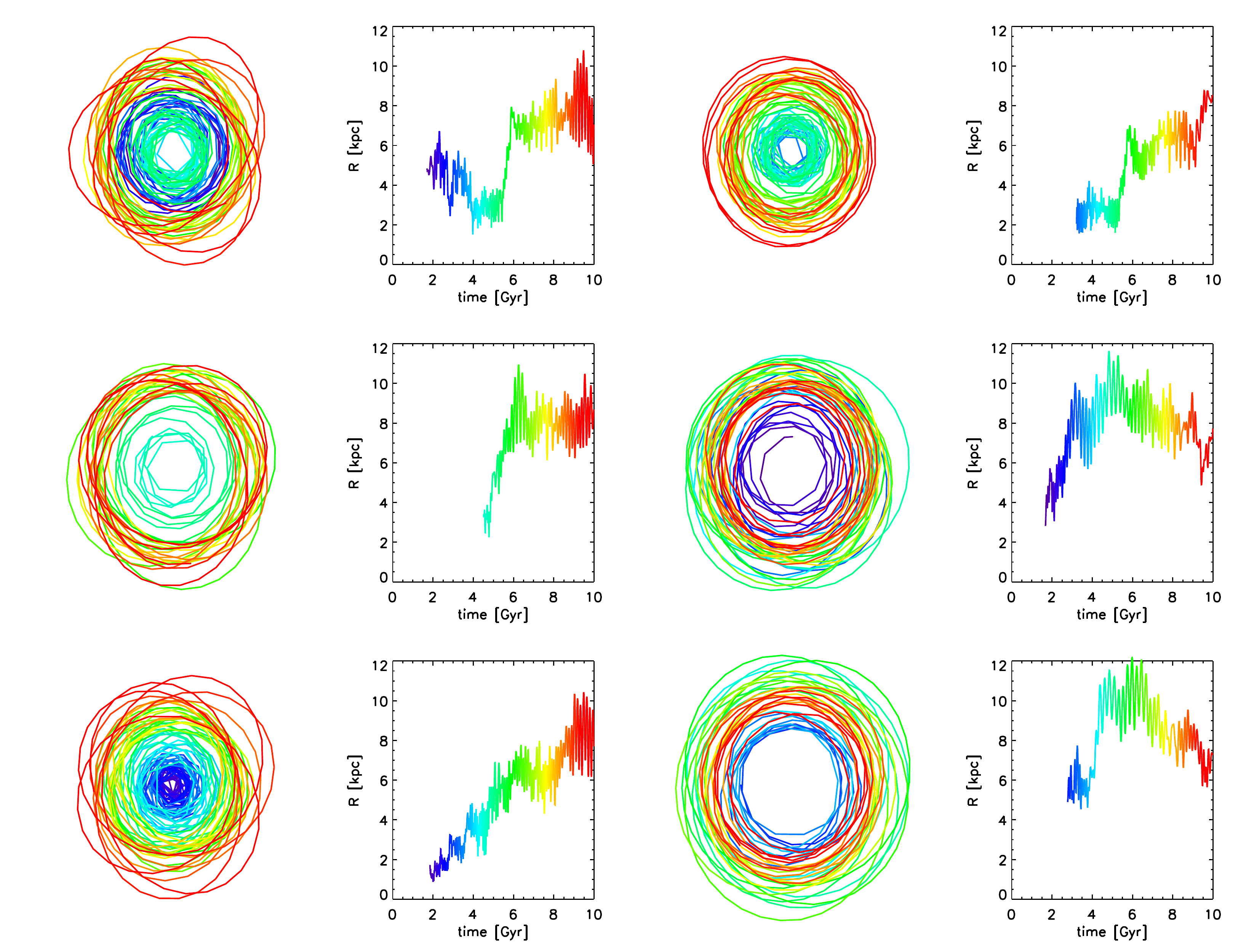}
\caption{Sample orbits from an idealized disk-formation simulation
  \citep{Roskar:2012}. All particles are selected to be between 7-9 kpc
  at the end of the simulation.}
\label{fig:orbits}
\end{figure*}

The expectation from \citet{Sellwood:2002} is therefore that if
transient spirals are present, migration \emph{must} take
place. Figure~\ref{fig:orbits} shows some examples of orbital
histories from an idealized disk-formation simulation from
\citet{Roskar:2012}. All of the stars were selected to be between 7-9
kpc at the end of the simulation, but clearly they originated in very
different places in the disk and had diverse orbital
histories. Interactions with spirals lead some stars to migrate
inwards, others migrate outwards, some orbits become more eccentric
and others circularize.

\begin{figure}
\centering
\includegraphics[width=4.in]{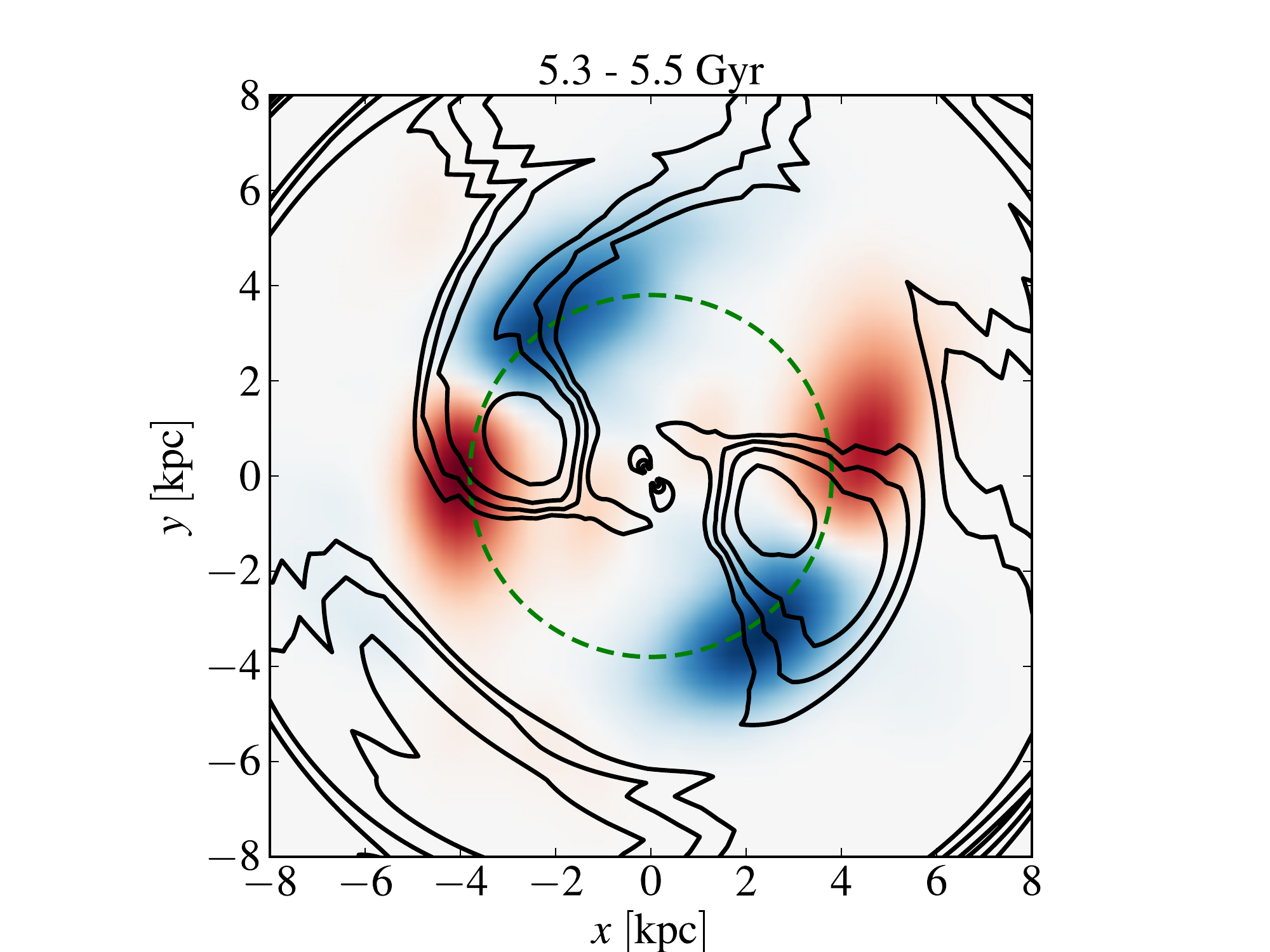}
\caption{Inward (red) and outward (blue) migrator density in the time
  interval from 5.3-5.5 Gyr. The particles are chosen only by virtue
  of being in the top 5 percent of the $\Delta L_z$ distribution over
  this time interval. The contours show the stellar density
  reconstructed from the m=2-4 Fourier components. The dashed green
  line shows the location of the corotation resonance for circular
  orbits.}
\label{fig:migrators}
\end{figure}

Investigating the reasons for migration in their models further,
\citet{Roskar:2012} confirmed that the most likely mechanism for the
large migrations occurring on short timescales was the corotation
resonance mechanism of
\citet{Sellwood:2002}. Figure~\ref{fig:migrators} shows one piece of
evidence supporting this hypothesis: the top 5\% of migrators in the
specified time interval straddle the peak of the spiral density wave
at the distance corresponding to its corotation resonance. In
addition, \citet{Roskar:2012} showed that the particles cross
corotation precisely at the time of peak spiral amplitude and that
their Jacobi integrals are conserved during the migration phase, all
of which are expected in the corotation resonance picture
\citep{Sellwood:2002}.

In the models of \citet{Sellwood:2002} and \citet{Roskar:2012}, the
spiral structure was shown, using Fourier methods
(e.g. \citealt{Sellwood:1986}), to host discrete pattern speeds with
well-defined resonances. However, in some models the spiral structure
pattern speed appears to match the rotation speed of the disk,
essentially making the entire disk co-rotate with the spiral
(e.g. \citealt{Grand:2012}). \citet{Kawata:2014} claim to have
detected a signature of such spiral structure in the MW through a
combination of measurements of gas and stellar motions. On the other
hand, \citet{Meidt:2009} used the modified Tremaine-Weinberg method to
recover pattern speeds in several external galaxies and found that in
general the pattern speeds were discrete, though the co-rotating
spirals tend to be found in disks with lower surface-densities. The
implications of co-rotating spirals is that mixing can be extremely
efficient, though it is not clear whether it should lead to a unique
chemo-dynamical signature.

Galactic disks are very prone to perturbations, so it is not
surprising that other suggestions for redistributing material in the
disks abound in the literature. For example, disks galaxies evolving
in a cosmological environment are expected to undergo frequent
encounters with substructure, which will not only disturb the disk
(e.g. \citealt{Kazantzidis:2009}) but also cause redistribution that
can mimic migration due to the corotation resonance
\citep{Quillen:2009}.

Another important consideration is the influence of bars on the
dynamical evolution of the disk. Bars are robust structures that once
formed evolve slowly. During the initial bar growth phase, angular
momentum is redistributed very efficiently but the process also
results in a significant amount of heating
\citep{Debattista:2006}. When fully formed, bars mostly heat the disk
through the inner and outer Lindblad resonances. Since bars are not
transient, efficient exchange of angular momentum is not expected at
the corotation resonance, because stars are mostly trapped on
horseshoe orbits. However, \citet{Minchev:2010} argued that in the
presence of overlapping resonances from other disk structure, the
horseshoe orbits can be disrupted and stars may efficiently gain or
lose angular momentum in a similar fashion to the transient spiral
mechanism of \citet{Sellwood:2002}. Such resonance overlap should
result in non-linear evolution near the resonances whenever multiple
patterns are present. \citet{Roskar:2012} looked for signatures of
this in their models where several spirals coexisted but found no
clear evidence of this process. The orbits of stars undergoing rapid
migration instead remained regular and clear signs of chaos were not
found. \citet{DiMatteo:2013} also argued that even steady bars can
drive migration well after the instability epoch, but it is not clear
whether the stellar dispersal in their model is due to heating and
disk spreading or genuine radial migration. The diffusion of ``hot
stars'' in barred galactic disks was shown already by
\citet{Pfenniger:1991}, but those stars are distinctly chaotic and
would result in non-circular orbits with large proper motions, which
is very different from the \citet{Sellwood:2002} migrators described
above.

In light of this discussion, it is crucial to point out that among the
various mechanisms that can move material around the disk quickly and
efficiently, migration due to the corotation resonance stands out
because it causes very little radial heating. Satellite perturbations
and bar formation, for example, affect the disk very differently and
these diverse processes should not all be placed under the common
label of ``radial migration''. They describe very different physical
processes and should be treated as such. However, just as the
corotation migration causes little heating, it also has the largest
effect on cool stellar populations. Therefore, models whose velocity
dispersions are too high may not show signatures of radial mixing that
are representative of the \citet{Sellwood:2002} migration
mechanism. Collisionless models with strong bars are strong candidates
for this category, since the bar formation episode heats the disk very
rapidly and can serve to prevent more subtle disk structure needed for
radial migration from forming.

\section{Influence of migration on disk stellar populations}
\subsection{Solar Neighborhood and Outer Disks}

\begin{figure}
\centering
\includegraphics[width=\textwidth]{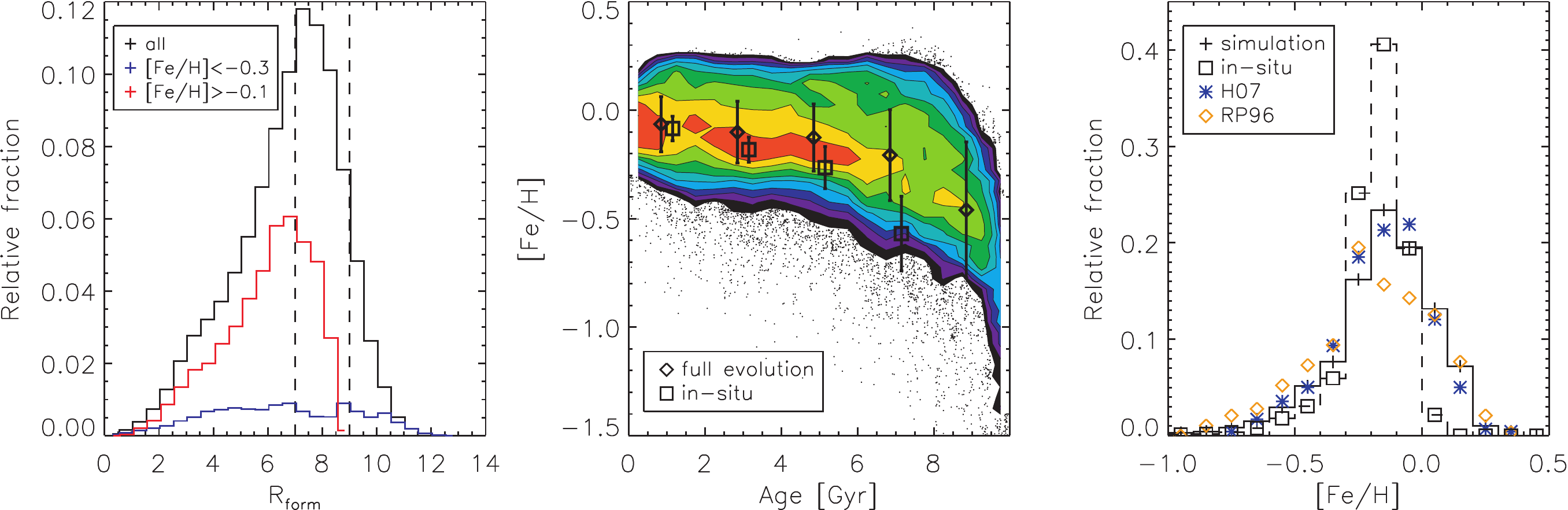}
\caption{\emph{Left:} distribution of formation radii for particles
  found within the solar neighborhood (defined to be $7<R<9$~kpc) at
  the end of the simulation from \citet{Roskar:2008a}. \emph{Center:}
  The age-metallicity relation (AMR) in the solar
  neighborhood. Migration adds significantly to the dispersion in
  metallicities at each age. \emph{Right:} Metallicity distribution
  function (MDF) is significantly broadened by migration.}
\label{fig:sn}
\end{figure}

One of the clear results of radial mixing is that it increases the
diversity of stellar populations everywhere in the disk, which is
especially important for Solar Neighborhood diagnostics like the AMR
relation \citep{Sellwood:2002,Roskar:2008,Schoenrich:2009}. The left
panel of Figure~\ref{fig:sn} shows the distribution of formation radii
for stars that end up between 7-9~kpc on mostly-circular orbits in the
simulation of \cite{Roskar:2008a}, clearly indicating that stars born
everywhere in the galaxy may find themselves in this part of the disk;
50\% of these stars originated at $R < 6$~kpc. Such redistribution has
obvious consequences for the AMR (middle panel) and the MDF (right
panel) -- the AMR is flattened and the dispersion in metallicity at
each age increases significantly, while the MDF is significantly
broadened.

A related and surprising early result of radial migration studies was
that in a truncated disk \citep[e.g.][]{Pohlen:2006}, the stellar age
profile is expected to show an upturn beyond the break radius
\citep{Roskar:2008}. This prediction has been confirmed by a number of
observational studies, both indirectly in terms of color gradients
\citep{Bakos:2008} and directly from stellar population modeling
using IFU observations \citep{Yoachim:2010} as well as HST
resolved-star data \citep{Radburn-Smith:2012}. While this prediction
is not unique to radial mixing, since it is possible that the outer
disks are influenced also by cosmological gas accretion and subsequent
star formation \citep[e.g.][]{Sanchez-Blazquez:2009}, radial migration
nevertheless can deposit significant amounts of material in outer
disks. The redistributed material can perhaps overshadow the stars
formed from accreted gas \citep{Roskar:2010} so the age inversion
remains one of the few robust predictions of the process that have
been confirmed by observations.

\subsection{Migrated stars in the thick disk?}

The thick disk is arguably one of the most coveted structures in the
Milky Way, because of its potential to inform us about the early epoch
of Milky Way's disk formation. The possibility that radial migration
can influence and partially populate the thick disk has therefore
received much attention in recent years. The concept is quite simple:
as a star changes its radial distance from the galactic center, the
change in the vertical restoring force felt at different parts of the
orbit affects its vertical displacement due to action conservation
\citep{Binney:2008, Schoenrich:2012}. This depends on how much of the
potential is due to the disk (the MW disk is believed to be close to
maximal, \citealt{Bovy:2013}) and the effect is compounded when the
star's guiding center is also changing.

In the model of \citet{Loebman:2011} stars far away from the mid-plane
at radii 7-9 kpc were found to be predominantly old and to originate
in the inner disk, compared with stars closer to the mid-plane which
were found to be younger and formed locally.  They also showed that
the $V_{\phi}$ - [Fe/H] were anti-correlated for young stars and only
slightly correlated for old stars in the solar neighborhood. This is
easy to understand in the context of migration because young stars
have not had time to migrate significantly and were therefore
populating the local volume by virtue of mild heating. The young stars
coming from the outer disk therefore had lower metallicities and
higher rotational velocities, while the opposite was true for those
coming from the inner disk. The old stars, on the other hand, had time
to migrate and mix and therefore showed only a weak trend between
metallicity and rotational velocity. The models qualitatively agree
with recent observations such as from \citet{Lee:2011}, which also
show a strong negative correlation between metallicity and $V_{\phi}$
for young ($\alpha$-deficient) stars and a slight positive correlation
for old ($\alpha$-enhanced) stars. Quantitatively, the models do not
agree with the data, though they also are not designed to reproduce
the MW. The flattening of the $V_{\phi}$-[Fe/H] relation for old stars
in the model is certainly due to the radial mixing but in this
particular model it does not turn into an overwhelmingly positive
correlation so it is not clear whether an external process is
required. Note that this observation poses a problem for all scenarios
where the disk grows from the inside out because it requires the metal
poor and metal rich stars to swap places in the disk at some
well-defined epoch, when the $\alpha$-enhancement is still high.

The influence of radial migration on the thickness of stellar
populations in the models from \citet{Loebman:2011} was explored in
detail by \cite{Roskar:2013}. They showed that vertical thickening is
a function of both age (i.e. heating) and radial displacement in the
disk. In other words, orbits of stars of the same age, which
experienced a similar amount of heating, become vertically more
extended if they migrate outwards and more compressed if they migrate
inwards. At the same time, stars from the same part of the disk that
migrate by a similar amount are thicker if they are older. These
results also showed that as stars move radially in the disk, their
vertical velocity dispersions decreased for inward migration and
increased for outward migration, confirming the effect shown by
\citet{Minchev:2012}.

It should be stressed that the increase of scale-height with outward
migration is not incompatible with the idea that the vertical action
is conserved and a consequent decrease in velocity dispersion. If the
velocity dispersion decreases as $\sigma_z \propto e^{-R/2R_d}$
\citep{Minchev:2012}, and we make an assumption that the disk is in
hydrostatic equilibrium with a sech$^2$ vertical density profile, it
is easy to show that the scale height $h_z \propto e^{R/R_d}$ and
therefore
\begin{equation}
\frac{h_{z,f}}{h_{z,i}} \propto \mathrm{exp}\left(\frac{\Delta R}{2R_d}\right),
\label{eq:hz}
\end{equation}
where $h_{z,i}$ and $h_{z,f}$ are the initial and final scale heights of
the population and $\Delta R$ is the change in radius. Thus, the
change in scale height with radial displacement follows naturally from
action conservation (see also \citealt{Schoenrich:2012} for a better
approximation).

\begin{figure}
\centering
\includegraphics[width=4in]{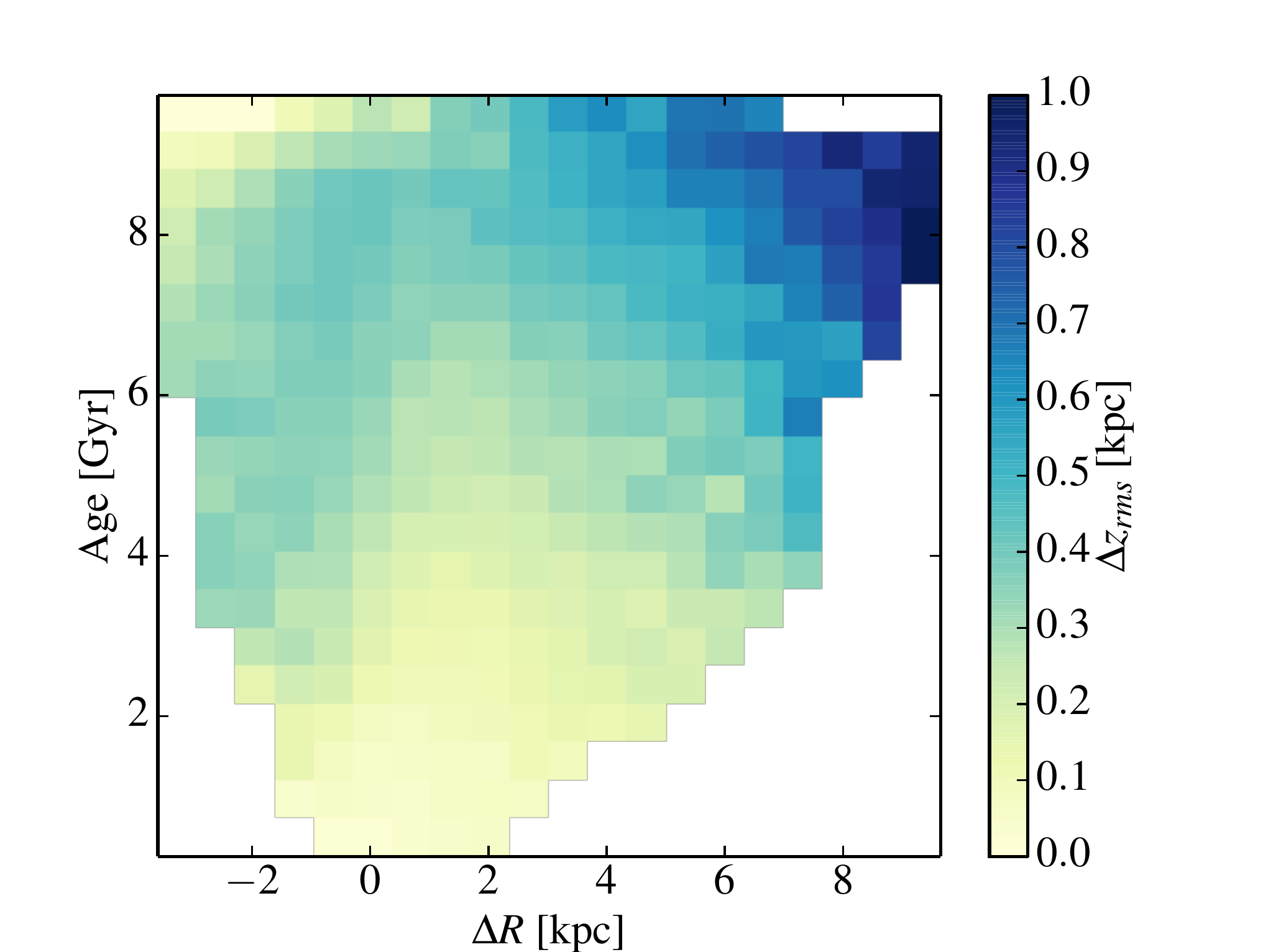}
\caption{Total change in $z_{rms}$ (proxy for thickness) for stars
  forming between 2-4 kpc as a function of age and change in radius.}
\label{fig:agedr}
\end{figure}

\subsection{Limitations of simulations}

The study of \cite{Roskar:2013} showed unequivocally that a single
population migrating outwards in their model will thicken, as expected
based on the fact that the vertical actions are, on average, conserved
\citep{Solway:2012}. Does this mean that migration can form a thick
disk? There are several caveats one must consider when interpreting
simulated disks and their usefulness in generalizing disk thickening.

First, a disk must be resolved vertically in order for the simulation
to say anything useful about the disk thickening process. In
\citet{Roskar:2013} the spatial resolution is 50~pc, but resolutions
of $>100$~pc are not uncommon in the literature dealing with vertical
disk evolution. This means that within approximately one scale-height
of the mid-plane, the forces on the particles are too low. The effect
of this on migration studies has not been carefully scrutinized, but
it is clear that it can affect how thick a migrated population looks
compared to the stars already present at a given radius.

Second, when investigating the effect of migrators on the in-situ
population, one must carefully consider the realism of the simulated
star-forming layer, or in the case of collisionless simulations the
assumptions about the initial disk structure. In \cite{Roskar:2013},
the gas disk out of which stars formed was significantly flared,
meaning that although stars from the inner disk increased their
thickness as they migrated outwards, they did not appear thicker than
the in-situ population \emph{because the in-situ population was
  thicker from the outset}.  This is a general problem with
hydrodynamic simulations using a temperature floor, since this means a
fixed dispersion at all radii and consequently a radially increasing
scale height $h_z$ for the young stars because $h_z \propto
\sigma_z/\Sigma$, where $\Sigma$ is the disk surface density. However,
while in that model the migrated stars therefore did not alone cause
the thickening of the disk, they themselves still thickened as they
migrated outwards. Therefore, they would not have been in the thicker
part of the disk in the first place was it not for the
migration. These stars therefore certainly influenced the chemical
abundance distributions at large distances from the plane even if
their effect on thickening was not large. The question of migration
and the thick disk is not necessarily whether migration can
\emph{create} a thick disk, but whether the chemical abundance
patterns that characterize the thick disk can be influenced by
migration.

Finally, numerical relaxation effects may have a disproportionate
influence on the vertical structure of disks
\citep{Sellwood:2013}. Therefore, numerical convergence tests must be
performed to determine the robustness of simulated structures, since
no obvious rule exists regarding sufficient particle numbers. This is
further complicated by the fact that the evolution of disks is often
highly stochastic, particularly when the disk supports multiple spiral
modes \citep{Sellwood:2009,Roskar:2012}. Nevertheless, such tests can
reveal strong divergences for certain parameter choices: for example,
\citet{Roskar:2012} showed that if they used 500~pc softening instead
of 100~pc or less, the disk formed completely different
non-axisymmetric structure. Determining whether numerical relaxation
is affecting the disk may be considerably more difficult for most
applications.

\section{Conclusions and challenges for the future}
By enriching the stellar diversity in any given part of the Galactic
disk, radial migration of stars complicates our interpretations of
the stellar population record left behind by the process of disk
formation. Migration due to the corotation resonance is inherently a
complicated process, due to its dependence on resonances with
transient structure that is difficult to constrain. Such migration is
special among the processes that redistribute material in galactic
disks, because it introduces very little heating and therefore leaves
behind no straightforward kinematic signatures.  For this reason,
$N$-body simulations have been instrumental in furthering our
understanding of the underlying process.

At the same time, the simulations exploring the observational effects
of migration have been very limited. On top of the dynamical
complexity, simulations try to add in the processes of star formation
and chemical enrichment in order to shed light on the expected effects
in stellar population trends that migration leaves behind. While the
purely dynamical modeling is potentially prone to numerical effects
(as discussed above in relation to resolving disk scale heights),
those are small compared to the uncertainties involved with sub-grid
modeling required for capturing star formation and chemical enrichment
within 3D $N$-body numerical simulations. We should therefore consider
it a huge triumph that these simulations can agree with observations
at all!

In the upcoming years, as data from Gaia, the GALAH survey
\citep{Freeman:2012}, and other similar large-scale projects become
available, simulations will be critical in aiding our interpretations
of those results. Nevertheless, to fully utilize those datasets, the
theoretical efforts need to strive for an understanding of the
leverage that various numerical and physical parameters exert on the
process of radial mixing. On the other hand, one hopes that chemical
tagging \citep{Freeman:2002, Mitschang:2013} will directly constrain
how much radial mixing has taken place in the MW disk. Combining such
constraints with a clearer picture of the underlying physical drivers
of the mixing process, we should be in a much better position to
unravel the past history of our MW disk.

\bigskip
\noindent {\bf {\large Acknowledgments}}\\
RR is funded in part by a Marie Curie Career Integration Grant. VPD is
supported by STFC Consolidated grant \#~ST/J001341/1.

\bibliographystyle{mn2e}

\begin{spacing}{1.0}
\bibliography{roskar_migration}
\end{spacing}

\end{document}